\documentclass[aps,superscriptaddress]{revtex4-2}
\usepackage[colorlinks=true, pdfstartview=FitV, linkcolor=blue, citecolor=red, urlcolor=magenta, breaklinks=true]{hyperref}
\usepackage{graphicx}  
\usepackage{amsfonts}
\usepackage{graphics}
\usepackage{color}
\usepackage{epsfig}
\usepackage{amssymb}
\usepackage{amsmath}
\usepackage{mathrsfs}
\usepackage{comment}

\begin{document}
\title{Thin-shell Gravastar Model in a BTZ geometry with minimum length}
\author{M. A. Anacleto}
\email{anacleto@df.ufcg.edu.br}
\affiliation{Departamento de F\'{\i}sica, Universidade Federal de Campina Grande
Caixa Postal 10071, 58429-900 Campina Grande, Para\'{\i}ba, Brazil}
\affiliation{Unidade Acad\^emica de Matem\'atica, Universidade Federal de Campina Grande
\\
58429-900 Campina Grande, Para\'{\i}ba, Brazil}
\author{A.~T.~N.~Silva}
\email{andersont.nsilva@gmail.com}
\author{L. Casarini}
\email{lcasarini@academico.ufs.br}
\affiliation{Departamento de F\'isica, Universidade Federal de Sergipe
49100-000 Aracaju, Sergipe, Brazil}

\begin{abstract} 
In this paper, we construct two spherically symmetric thin-shell gravastar models within a BTZ geometry with minimum length. Therefore, in the inner region of the gravastar, we consider an anti-de Sitter metric with minimum length. Thus, for the first model, we introduce the minimum length effect using the probability density of the ground state of the hydrogen atom in two dimensions. For the second gravastar model, we adopt a Lorentzian-type distribution. Also in the outer region, we consider the BTZ black hole metric. So, by examining the inner spacetime, the thin shell, and the outer spacetime, we find that there are different physical characteristics regarding their energy densities and pressures that make the gravastar stable. This effect persists even when the cosmological constant is zero. In addition, we determined the entropy of the gravastar thin shell. Besides, we explore the thermodynamic properties of the BTZ black hole with minimum length in Schwarzschild-type form and also check its stability.

\end{abstract}
\maketitle
\pretolerance10000
\section{Introduction}
In recent years, both cosmology and astrophysics have been the subject of scientific interest in the quest to explore fundamental questions surrounding the nature of the universe. Advances in observations such as the shadows of black holes in the core of the galaxy M87 \cite{EventHorizonTelescope:2019dse,EventHorizonTelescope:2021bee} and the neighboring galaxy Sgr $A^{*}$ \cite{Gillessen:2008qv}, the radiation emitted by accretion disks in active galactic nuclei \cite{Arsioli:2024zwq}, as well as the direct detection of gravitational waves resulting from the merger of binary black hole systems, as recorded by LIGO/Virgo \cite{KAGRA:2021duu}, and the creation of new theoretical frameworks, contribute to a deeper understanding of cosmic structures and astrophysical processes that shape our universe. This continued focus stems from discoveries that challenge conventional models and pave the way for new models and concepts that are not yet fully understood.
{For this reason, compact objects are a crucial source, as they provide the necessary conditions to test many pertinent ideas in the high-density domain}. One of the most interesting and challenging problems in modern astrophysics involves these compact astrophysical objects. Black holes represent the endpoint of the collapse of massive stars, which can be described by Einstein's theory of relativity, and the evidence of their existence can be confirmed through the observation of gravitational waves \cite{LIGOScientific:2016vlm}. This is because the catastrophic event that triggered the wave carries the expected characteristics of a binary black hole system in the merger. 

However, other possibilities remain to explain the final fate of gravitational collapse. Because of this, a large body of literature on compact astrophysical objects with properties similar to those of black holes has emerged in recent decades (see, for example, \cite{Cardoso:2019rvt}  for a review). It is within this extensive body of work that the ingenious solution proposed by Mazur and Mottola ~\cite{Mazur:2001fv,Mazur:2004fk} in 2001 emerges as an alternative model to black holes as the endpoint of gravitational collapse, named the gravastar model (gravitational vacuum star), as an alternative to the Schwarzschild solution \cite{Schwarzschild:1916uq}. According to their model, the gravastar in particular has three separate zones with different equations of state (EoS): An inner region filled with dark energy with an isotropic de Sitter vacuum situation ($p = -\rho$), a thin-shell intermediate layer consists of rigid fluid matter ($p = \rho$)and the outer space is empty, and Schwarzschild geometry ($p=\rho = 0$) represents this situation appropriately.
The general idea is to prevent the formation of event horizons (and singularities) so as not to allow the collapse of matter at or near the event horizon. In other words, another black hole structure could be formed by the gravitational collapse of a massive star.

Recent studies of the brightness of distant type Ia supernovas~\cite{SupernovaSearchTeam:1998fmf, SupernovaCosmologyProject:1998vns, Bahcall:1999xn, Planck:2018vyg} suggest that the expansion of the universe is faster than previously believed. This suggests that cosmic pressure $p$ and energy density $\rho$ must contradict the strong energy condition, that is, $ \rho + 3p < 0$. The so-called ``Dark energy'' is the component that allows this demand to be met at a specific stage of cosmic evolution \cite{Sahni:1999gb, Peebles:2002gy, Padmanabhan:2002ji}. Several substances determine the dark energy condition. The best-known proposal involves a non-vanishing cosmological constant, which is equivalent to the fluid that satisfies the EOS $p = - \rho$ \cite{Pradhan:2023wac}. 

In~\cite{Usmani:2010ac}, Usmani et al. introduced an innovative approach by developing a new model of a gravastar, assuming motion, with a charged interior and, on its exterior, the line element for this approach was Reissner-Nordstr\"{o}m, instead of the conventional Schwarzschild model. Subsequently, Rahaman et al.~\cite{Rahaman:2011we} designed a spherically symmetric neutral model of a gravastar in (2+1) anti-de Sitter spacetime, expanding the scope beyond studies performed in (3+1) dimensions. 
The external configuration of the gravastar model by Rahaman et al.~\cite{Rahaman:2011we} corresponds to the (2+1) anti-de Sitter exterior spacetime, reminiscent of the BTZ black holes, as proposed by Bañados, Teitelboim, and Zanelli~\cite{Banados:1992wn}. 
Furthermore, the study~\cite{silva2024} analyzes essential properties of black holes, such as Hawking temperature, entropy, and heat capacity. In its limit, spacetime adopts an anti-de Sitter configuration, thus exhibiting a conformal structure that is especially significant for studies within the framework of the AdS/CFT correspondence~\cite{DalBoscoFontana:2023syy}. 
These investigations are conducted in a controlled manner, enabling the analytical investigation of the impacts of quantum corrections. Thus, the results obtained offer solid insights into how the minimum length relates to the stability and thermodynamics of compact and exotic objects, representing a natural and logical intermediate step before generalizing to (3+1) dimensions. 

The motivation for employing minimum-length BTZ geometry in modeling a thin-shell gravastar is linked both to the search for physically consistent options regarding complete gravitational collapse and to the requirement to include quantum effects in macroscopic descriptions of gravity. In this scenario, the gravastar presents itself as an alternative that prevents the creation of event horizons and singularities, replacing the infinite-density core with an internal region dominated by de Sitter vacuum-type energy. The use of a (2+1)-dimensional scenario, typical of BTZ geometry, offers a theoretically controlled environment that allows for the analytical analysis of the effect of quantum corrections before proceeding to more realistic generalizations in (3+1) dimensions. This proves especially beneficial in research related to spacetimes with a negative cosmological constant. In this context, the implementation of a minimum length is fundamental, functioning as a physical regulator that eliminates Hawking temperature divergence and prevents the appearance of non-physical thermodynamic behaviors, such as the entropy of the divergent shell in the limit where this parameter approaches zero. Furthermore, this minimum length directly affects the stability of the configuration. In models with a Lorentzian distribution, it can act as a cosmological constant that sustains the formation of the gravastar, even without a fundamental cosmological constant. Thus, the inclusion of a minimum length not only alters the geometric configuration of spacetime but also enables the analysis of how quantum effects linked to the transition between the inner and outer regions affect the thermodynamic properties and stability criteria of the thin shell.
By introducing noncommutativity in spacetime as an adjustment of the metric of the inner region of a gravastar, through a Lorentzian distribution~\cite{Anacleto:2019tdj,Zeng:2021dlj,Zeng:2022fdm,Campos:2021sff,Anacleto:2020efy,Anacleto:2020zfh,Hu:2023eow,Anacleto:2022shk,Saleem:2023pyx,AraujoFilho:2024rss,Hamil:2024ppj,Wang:2024fiz,Jha:2022bpv,Jha:2023htn}, to keep it stable, Silva et al.~\cite{silva2024} showed that the parameter $\theta$ works as an effective instrument in the absence of the cosmological constant. Furthermore, some work on gravastar has been explored  in noncommutative geometries\cite{BANERJEE2016,Lobo:2010uc,Das:2018fzc,Ovgun:2017jzt}, modified gravity theories~\cite{Yousaf:2019zcb,Shamir:2018qhq,Shamir:2020apc,BHAR2021}, rainbow gravity~\cite{Barzegar:2023ueo}, and also in other models~\cite{Rosa:2024bqv,Khlopov:1985jw,Konoplich:1999,Khlopov:2000,Khlopov:2008qy}. 

As established by the famous Bekenstein-Hawking relation, in gravitational systems, entropy is strongly associated with the enveloping surfaces~\cite{Bodendorfer:2014fua}. In the case of black holes, the entropy of an event horizon is directly proportional to its area. Although gravastars do not have an event horizon like black holes, the thin shell that delimits the boundary between the inner and outer regions plays a similar role in the calculation of entropy.
The inner surface of gravastars can be interpreted as an area where quantum fluctuations occur. Under these circumstances, the entropy in the layer symbolizes the amount of information “hidden” by the transition between the inner and outer regions. Since the BTZ metric with minimum length changes the spatial structure, the entropy needs to take these configurations into account. Therefore, when we check the entropy in the thin layer of the gravastars presented here, we are also checking how the minimum length in their interior affects their stability. 

In this work, we will focus on a type of gravastar in which we will consider a BTZ metric with minimum length in its inner region and a geometry associated with a BTZ solution in the outer region, both united, at their limits, by a thin shell. Thus, we will verify the stability conditions based on its surface energy density and surface pressure. Then, we will show that the stability conditions are satisfied even with the zero cosmological constant. Also, we will determine the entropy in the thin shell. In addition, we will perform a thermodynamic analysis of the BTZ black hole with minimum length in Schwarzschild-type form in three-dimensional spacetime. 

The paper is organized into distinct sections and subsections. In Sec.~\ref{BTZML}, we introduce noncommutative through a minimum length and analyze these effects in the calculation of Hawking temperature, entropy, and the specific heat capacity. In Sec.~\ref{stbtzgrav}, we apply the probability density of the ground state of the hydrogen atom to determine the energy density and pressure at the surface, formulate the structural equations of the gravastar, examine the matching conditions at the interface, and we check the entropy in the thin shell. In Sec.~\ref{LTDist}, we adopt a Lorentzian-type distribution to examine the energy density and pressure at the surface and check the entropy in the thin shell. In Sec.~\ref{conc}, we make our final considerations.
\newpage
\section{BTZ BLACK HOLE WITH MINIMUM LENGTH} \label{BTZML}
In this section, we explore the effect of the minimum length on the calculation of the Hawking temperature, entropy and heat capacity of the BTZ black hole in Schwarzschild-type form.

\subsection{Probability Density}
Here, we introduce the minimum length contribution into the BTZ metric by modifying the mass density as follows~\cite{Yang1991,Miao:2016ipk,Anacleto:2022sim}:
\begin{eqnarray}
\rho(r) = \frac{M_0}{\gamma^{2}\pi} \exp\left(\frac{-4r}{\gamma}\right),
\label{dens}
\end{eqnarray}
being $M_0$, the total mass spread over the entire linear-sized region $\sqrt{\gamma}$ of the BTZ black hole and $\gamma$, is the minimal length. 
In this case, the ``stained'' mass is distributed as follows~\cite{Anacleto:2022sim}:
\begin{eqnarray}
\mathcal{M} = \int_{0}^{r} \rho(r)2 \pi r dr = M_0 \left[1 - \frac{(4r + \gamma)}{\gamma} \exp\left(\frac{-4r}{\gamma}\right) \right]. 	
\end{eqnarray} 
Thus, we have the line element given by
\begin{eqnarray}
ds^{2} =  -f(r)dt^{2} + f(r)^{-1}dr^{2} + r^{2}d\phi^{2},
\label{btzcm0}
\end{eqnarray}
where the metric function is given by
\begin{eqnarray}
f(r)
 = -M_{0} + \left[\frac{(8M_{0}r+2\gamma)}{\gamma}\exp\left(\frac{-4r}{\gamma}\right)\right] +\frac{r^2}{l^2}.
\label{fbtzcm}
\end{eqnarray}
For $r/\gamma \rightarrow\infty$, the BTZ black hole metric is restored.

The radius of the event horizon for metric (\ref{fbtzcm}) is given by
\begin{eqnarray}
r_H=r_h\left[1 - \left(\frac{l^2}{r^2_h} + \frac{4 r_h}{\gamma} \right)\exp\left(\frac{-4r_h}{\gamma}\right)  \right] + \mathcal{O}\left(e^{{-4r_h}/{\gamma}}\right)^2,    
\end{eqnarray}
being $r_h=\sqrt{M_0l^2}$, the event horizon for the nonrotating BTZ black hole. 

\begin{figure}[!htb]	
\centering
\includegraphics[scale=.28]{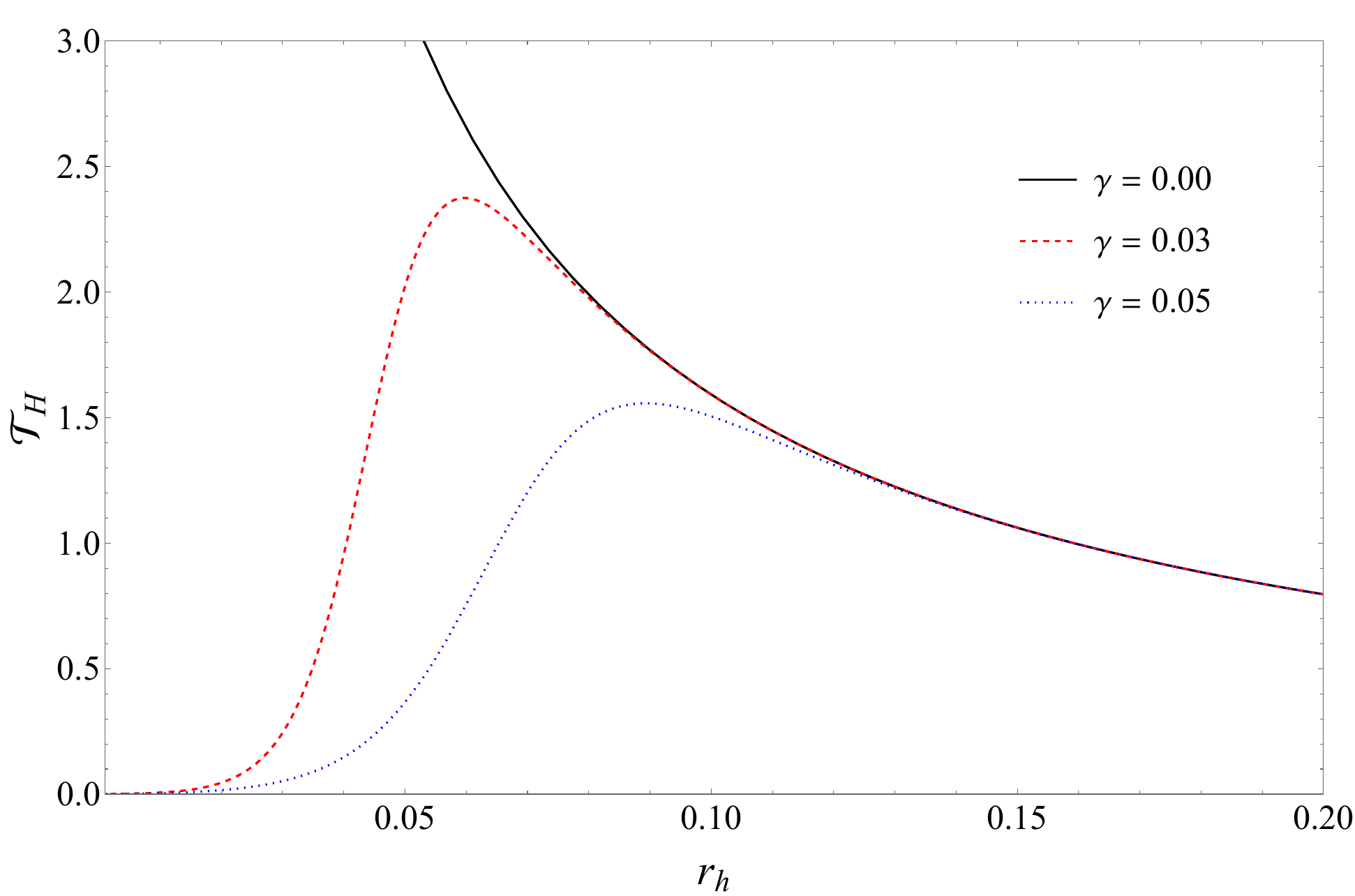}
\caption{\footnotesize{Hawking temperature as a function of the horizon radius $r_{h}$ for  $\gamma = 0$ (black), $\gamma = 0.03$ (red, dashed) and $\gamma = 0.05$ (blue, dotted), assuming $ l=1 $. The $\gamma$ value was chosen because there is a minimal length requirement with small parameter values. }
}
\label{fg0}
\end{figure}

Hence, we obtain the corrected Hawking temperature due to the minimum length as follows~\cite{Carlip:2005zn,Anacleto:2020efy,silva2024}:
\begin{eqnarray}
T_H=\frac{f^{\prime}(r_H)}{4\pi}= \frac{r_h}{2\pi l^2 } - \frac{r_h}{2\pi l^2}  \left[  \frac{l^2}{r^2_h} + \frac{4 r_h}{\gamma} + \frac{16 r^2_h}{\gamma^2} \right]\exp\left(\frac{-4r_h}{\gamma}\right).
\end{eqnarray}
This definition has been used for comparison with many works in the literature.
The result can be expressed in Schwarzschild-type form, given by:
\begin{eqnarray}
\mathcal{T}_H=\frac{T_H}{M_0}
= \dfrac{1}{2\pi\left[ r_h + \left(\dfrac{l^2}{r_h} + \dfrac{4 r^2_h}{\gamma} + \dfrac{16 r^3_h}{\gamma^2}\right) 
\exp\left(\dfrac{-4r_h}{\gamma}\right)\right]}.
\label{TH}
\end{eqnarray}
Note that the Hawking temperature reaches a maximum point before going to zero when the horizon radius, $r_h$,
tends to zero, as shown in Fig.~\ref{fg0}.

Then, by applying the condition $f(r)=0$, we obtain
\begin{eqnarray}
\Tilde{M}=\frac{r^2_H}{l^2}\left[1-2\left( \frac{l^2}{r^2_h}+\frac{4r_H}{\gamma}\right)\exp\left(-\frac{4r_H}{\gamma}\right)\right]^{-1}=\frac{r^2_h}{l^2} + \left(e^{-4r_h/\gamma}\right)^{2}.
\label{massa}
\end{eqnarray}

Next, we determine the entropy of the BTZ black hole with minimum length using the following relation
\begin{eqnarray}
S=\int \frac{1}{\mathcal{T}_{H}}\frac{\partial \Tilde{M}}{\partial{r}_h} d{r}_h, \qquad 
\frac{\partial \Tilde{M}}{\partial{r}_h}=\frac{2r_h}{l^2}=\frac{2M_0}{r_h}.
\label{ent}
\end{eqnarray}

By replacing (\ref{TH}) in (\ref{ent}), we find
\begin{eqnarray}
S=\int {2\pi r_h}\left[1+ \left(\frac{l^2}{r^2_h}+\frac{4r_h}{\gamma} +\frac{16r^2_h}{\gamma^2}\right)e^{-4r_h/\gamma} + \cdots\right]\left(\frac{2M_0}{r_h} \right) dr_h.
\end{eqnarray}
Hence, we obtain
\begin{eqnarray}
\mathcal{S}=\frac{S}{M_0}=4\pi r_h  -4\pi\left[\frac{3 \gamma}{4} +\frac{\gamma l^2}{4r^2_h} + 3r_h +\frac{4r^2_h}{\gamma} \right]e^{-4r_h/\gamma} +\cdots .
\label{entrexp}
\end{eqnarray}

Now, to compute the specific heat capacity, we use the following relation:
\begin{eqnarray}
C=\frac{\partial \tilde{M}}{\partial \mathcal{T}_H}=\frac{\partial \tilde{M}}{\partial r_h}\left(\frac{\partial \mathcal{T}_H}{\partial r_h}\right)^{-1}.
\end{eqnarray}
Then, we find
\begin{eqnarray}
\label{shcg}
\mathcal{C}=\frac{C}{M_0}=-4\pi r_h \left[ 1 - \left(\frac{3l^2}{r^2_h} + \frac{4 l^2}{\gamma r_h} +\frac{64r^3_h}{\gamma^3}\right) e^{-4r_h/\gamma}\right].
\end{eqnarray}
For $\gamma\ll 1$ and considering the dominant term in the above result, we obtain
\begin{eqnarray}
\label{Cs}
 \mathcal{C}\approx -4\pi r_h \left[ 1 -\frac{64r^3_h}{\gamma^3} e^{-4r_h/ \gamma}\right].
\end{eqnarray}

\begin{figure}[!h]	
\centering
\includegraphics[scale=.35]{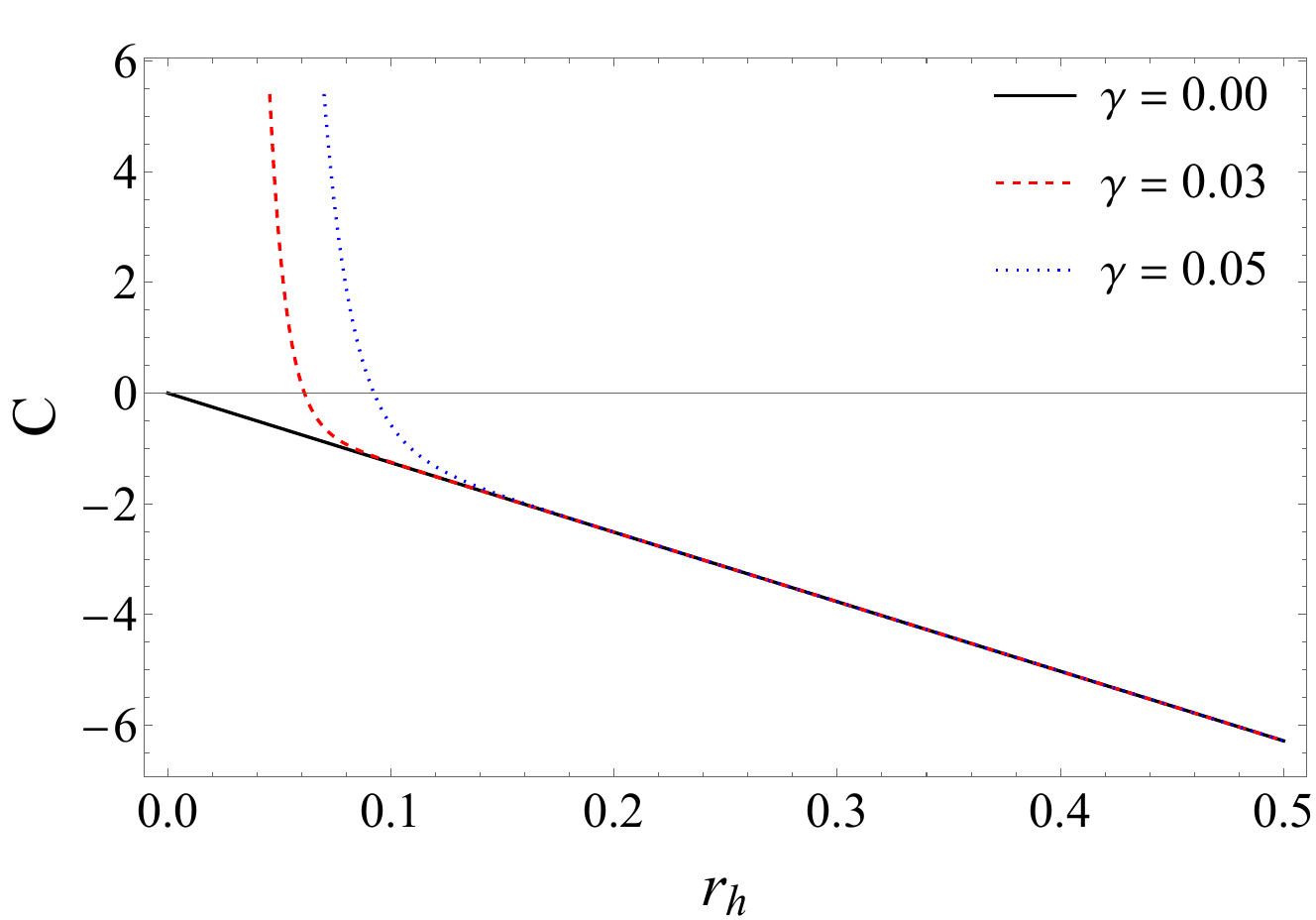}
\caption{\footnotesize{Specific heat capacity $\mathcal{C}$ as a function of the horizon radius $r_{h}$ for $\gamma=0$ (Black) $\gamma = 0.03$ (red, dashed) and $\gamma = 0.05$ (blue, dotted), assuming $ l=1 $.}}
\label{fg001}
\end{figure}
Therefore, the specific heat capacity becomes zero when the following condition is satisfied:
\begin{eqnarray}
\label{condcs}
\frac{ r^3_h}{ e^{4r_h/\gamma}}=\left(\frac{\gamma}{4}\right)^3.
\end{eqnarray}
Consequently, the black hole ceases to completely evaporate and becomes a remnant. 
In Fig.~\ref{fg001}, we show the behavior of the specific heat capacity. 
Then, it is seen that for $0< r_h <r_{min}$, the black hole enters a phase of stability with $\mathcal{C}>0$.

Thus, applying the condition (\ref{condcs}), the Hawking temperature takes the form
\begin{eqnarray}
\mathcal{T}_{H}=\frac{1}{2\pi \left( r_h + \dfrac{\gamma}{4} + \dfrac{\gamma^2}{16r_h}\right) }.
\label{tempm}
\end{eqnarray}

\begin{figure}[!htb]	
\centering
\includegraphics[scale=.28]{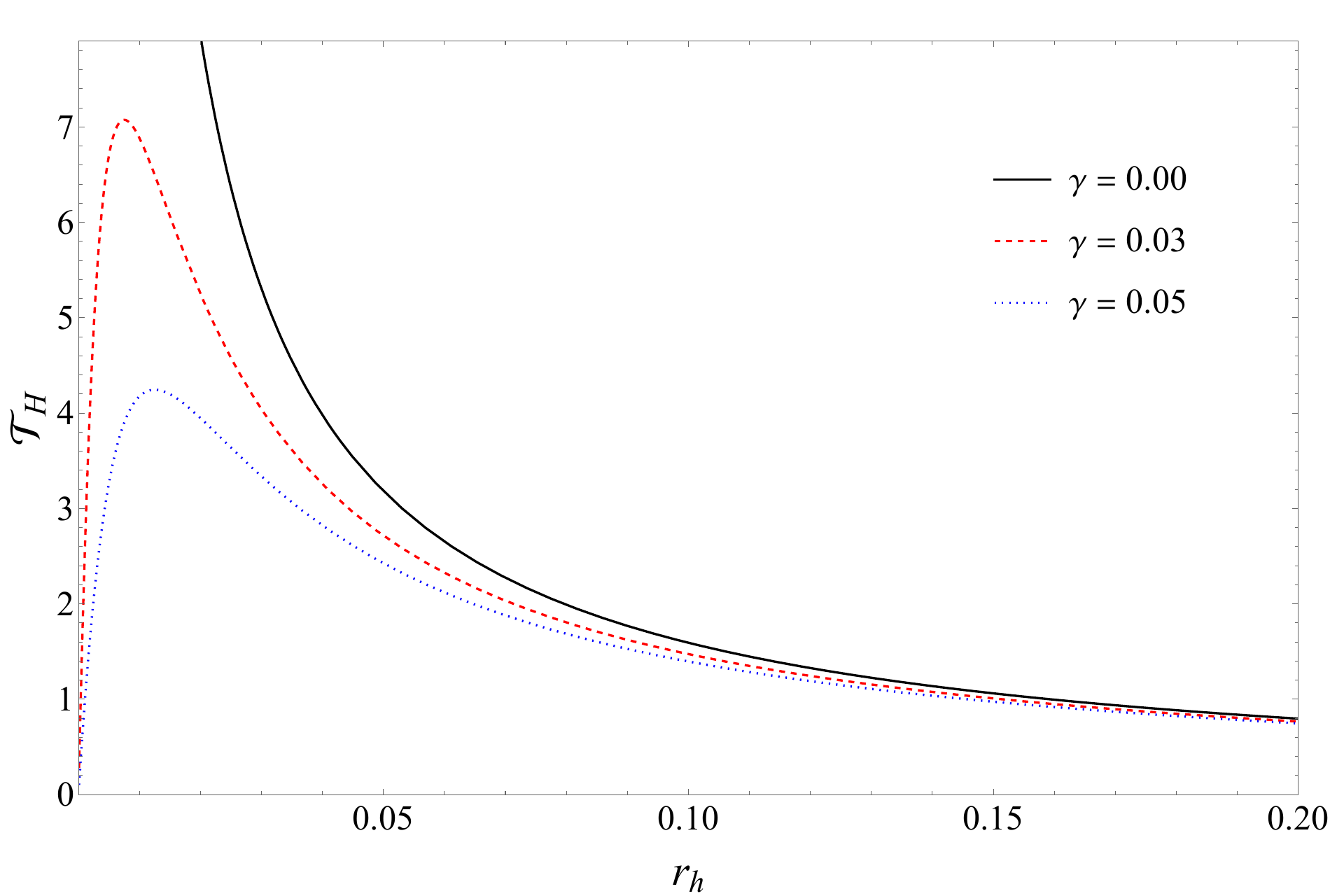}
\caption{\footnotesize{Hawking temperature as a function of the horizon radius $r_{h}$ for  $\gamma = 0$ (black), $\gamma = 0.03$ (red, dashed) and $\gamma = 0.05$ (blue, dotted). The $\gamma$ value was chosen because there is a minimal length requirement with small parameter values. }}
\label{figthexp}
\end{figure}

Hence, for $\gamma=0$, we recover the temperature of the BTZ black hole in Schwarzschild-like form.
From equation (\ref{condcs}), we find a minimum radius $r_{min}\approx\gamma/4$ and a minimum mass $M_{min}=r^2_{min}/l^2=\gamma^2/16l^2=-\Lambda\gamma^2/16$.
Then, by substituting into the above equation, we obtain a maximum temperature given by
\begin{eqnarray}
\mathcal{T}_{Hmax}=\frac{1}{6\pi r_{min}} =\frac{2}{3\pi\gamma}.  
\end{eqnarray}
Thus, for $\gamma=0.03$ and $\gamma=0.05$, we have the following maximum temperatures given respectively by: 
$\mathcal{T}_{Hmax}=7.07355$ and $\mathcal{T}_{Hmax}=4.24413$, which are in accordance with the graphs shown in Fig.~\ref{figthexp}.

Now, considering the Hawking temperature from equation (\ref{tempm}), we obtain the following result for the entropy
\begin{eqnarray}
\mathcal{S}=4\pi r_h + \pi\gamma\ln r_h - \frac{\pi\gamma^2}{4 r_h}.    
\end{eqnarray}
In this case, a logarithmic correction term is obtained for the entropy of the modified BTZ black hole.

\subsection{Lorentzian-Type Distribution}						
At this point, we will analyze the contribution of the minimum length considering a mass distribution of the form~\cite{Anacleto:2022sim}
\begin{equation}
\rho(r) = \dfrac{16 M_{0}\beta}{\pi(4r + \beta)^{3}},
\label{atn1}  
\end{equation}
where $\beta$ is the minimum length parameter. 
Hence, we find that the mass is given by
\begin{eqnarray}
\mathcal{M}_\beta(r) &=&   \int^{r}_{0} \rho(r)2\pi r dr = \frac{16M_{0}r^2}{(\beta + 4 r)^2},
\\
&=& M_{0} -\frac{M_{0} \beta}{2 r} + \frac{3M_{0} \beta^2}{16r^2} + {\cal O}(\beta^3).
 \label{m1}
\end{eqnarray}
Then the line element becomes~\cite{Anacleto:2022sim}
\begin{eqnarray}
ds^{2} = -g(r)dt^{2} + g(r)^{-1}dr^{2} + r^{2}d\phi^{2},
\label{metcs}
\end{eqnarray}
where the metric function is written as follows
\begin{eqnarray}
g(r)=-M_{0} +\frac{M_{0}\beta}{2r}+\frac{r^2}{l^2} - \frac{3M_{0} \beta^2}{16r^2} .
\label{fmetcs} 
\end{eqnarray}
Note that correction terms are induced in the meric function. The first correction term is of the Schwarzschild type, and the last correction term is related to the effective angular momentum contribution.

Hence, the horizons are given by~\cite{Anacleto:2022sim}
\begin{eqnarray}
&&r_{+}=r_h - \frac{\beta}{4} + \frac{3\beta^2}{32 r_h},
\\
&&r_{-}=\frac{\beta}{4} - \frac{3\beta^2}{32 r_h},
\end{eqnarray}
where, $r_h=\sqrt{l^2M_0}$ is the event horizon for the nonrotating BTZ black hole.

\begin{figure}[!htb]	
\centering
\includegraphics[scale=.4]{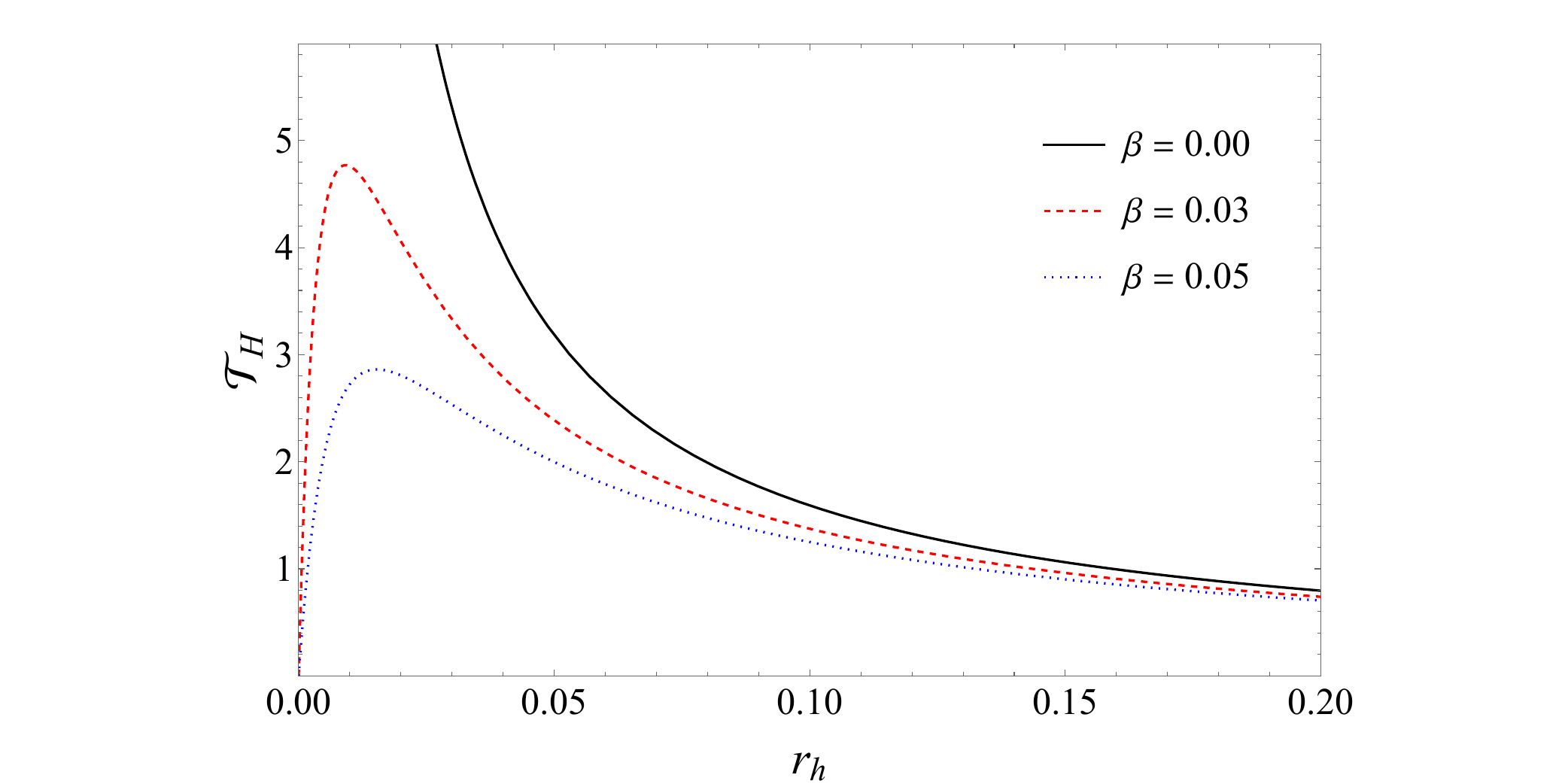}
\caption{\footnotesize{Hawking temperature as a function of the horizon radius $r_{h}$ for  $\beta = 0$ (black), $\beta = 0.03$ (red, dashed) and $\beta = 0.05$ (blue, dotted). The $\beta$ value was chosen because there is a minimal length requirement with small parameter values. }}
\label{figth}
\end{figure}

Hence, the Hawking temperature of the nonrotating BTZ black hole with minimum length reads~\cite{Carlip:2005zn}
\begin{eqnarray}
 {T}_H&=&\frac{g^{\prime}(r_{+})}{4\pi}
=\frac{r_{h}}{2\pi l^2 }\left[ 1 - \dfrac{\beta}{2 r_{h}} + \dfrac{5\beta^2}{32 r_{h}^{2}}\right].
\end{eqnarray}
The result can be expressed in Schwarzschild-type form as follows:
\begin{eqnarray}
\mathcal{T}_H=\frac{T_H}{M_0}=\dfrac{1}{2\pi\left[ r_h  + \dfrac{\beta}{2} + \dfrac{3\beta^2}{32 r_{h}} + \cdots\right]}.
\label{THL}
\end{eqnarray}
Note that the Hawking temperature reaches a maximum point before going to zero when the horizon radius, $r_h$,
tends to zero. 
Therefore, as we can see in Fig.~\ref{figth}, the minimum length parameter $\beta$ plays the role of a regulator removing the Hawking temperature singularity of the BTZ black hole in Schwarzschild-like form.

In order to compute the entropy of the BTZ black hole, we express the mass as
\begin{eqnarray}
\tilde{M}=\frac{r^2_h}{l^2} - \frac{\beta^2}{16l^2}+\cdots.
\end{eqnarray} 

Next we determine the entropy by applying the following equation
\begin{eqnarray}
S=\int \frac{1}{\mathcal{T}_{H}}\frac{\partial \Tilde{M}}{\partial{r}_h} d{r}_h, \qquad
\frac{\partial \Tilde{M}}{\partial{r}_h}= \frac{2r_h}{l^2}=\frac{2M_0}{r_h}. 
\label{ent1}
\end{eqnarray}
Then, for entropy we find
\begin{eqnarray}
\mathcal{S}=\frac{S}{M_0}= \int \frac{2}{r_h\mathcal{T}_{H}} d{r}_h=
4\pi r_h + \frac{\pi\beta}{2}\ln r_h -\frac{3\pi\beta^2}{32 r_h}.
\label{entrlo}
\end{eqnarray}
Therefore, a logarithmic correction term for the entropy has been obtained.

\begin{figure}[!h]	
\centering
\includegraphics[scale=.35]{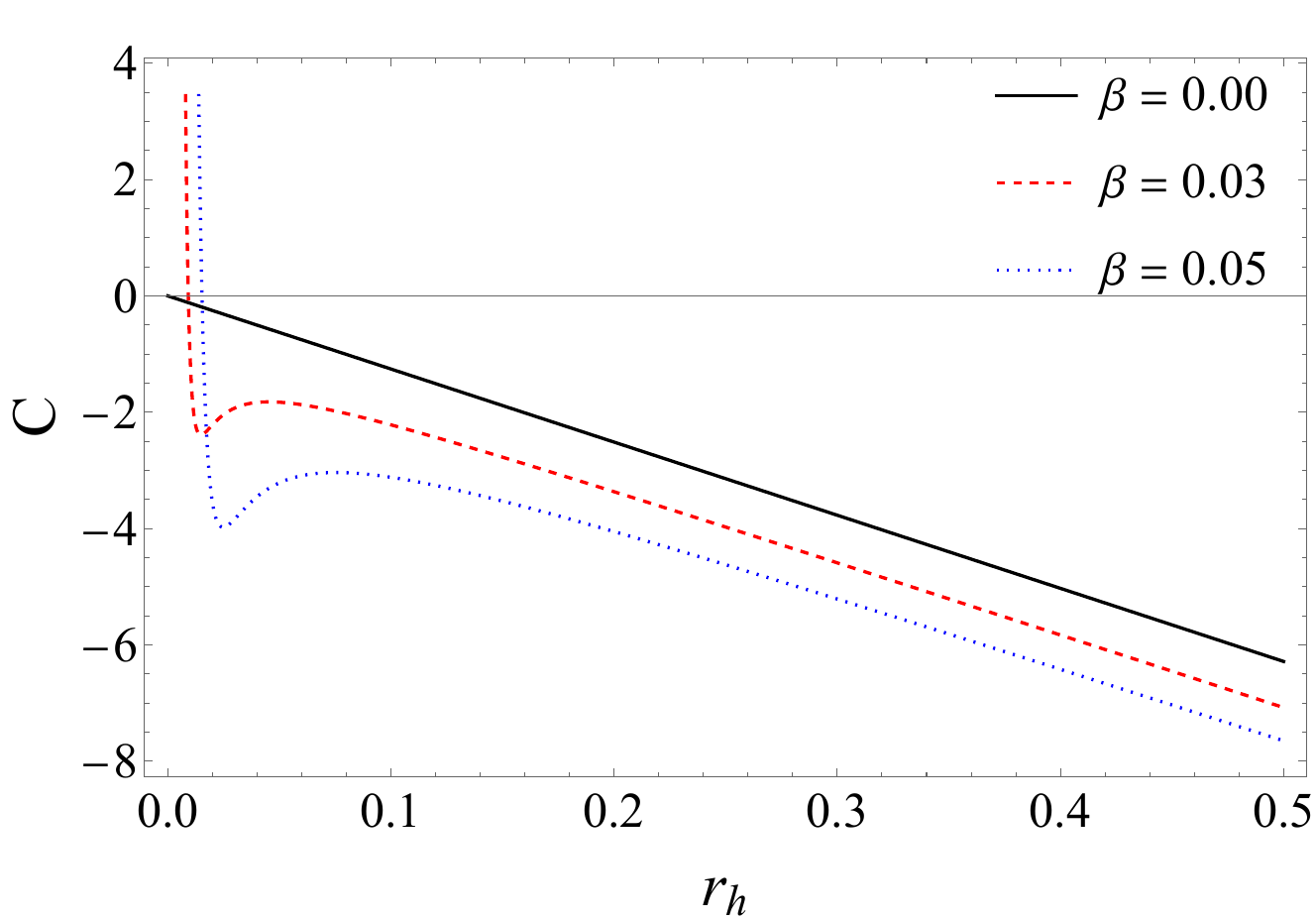}
\caption{\footnotesize{Specific heat capacity $\mathcal{C}$ as a function of the horizon radius $r_{h}$ for $\beta=0$ (Black), $\beta = 0.03$ (red, dashed) and $\beta = 0.05$ (blue, dotted).}}
\label{fg002}
\end{figure}

For the heat capacity, we have
\begin{eqnarray}
C=\frac{\partial \tilde{M}}{\partial \mathcal{T}_H}=\frac{\partial \tilde{M}}{\partial r_h}\left(\frac{\partial \mathcal{T}_H}{\partial r_h}\right)^{-1}.
\end{eqnarray}
Thus, we find the following correction for the specific heat capacity
\begin{eqnarray}
\mathcal{C}=\frac{C}{M_0}\approx -4 \pi r_h \left(1+\frac{\beta}{r_h} + \frac{6\beta^2}{32 r_h^2}\right)^2
\left(1 - \frac{3 \beta^2}{32 r^2_h}\right).
\label{caplo}
\end{eqnarray}
For $r_h=\dfrac{1}{4}\sqrt{\dfrac{3}{2}}\beta$, we have $\mathcal{C}=0$, and, thus, we find that the BTZ black hole in Schwarzschild-type form stops evaporating completely, becoming a remnant. 
This can be seen in Fig.~\ref{fg002}. 
Thus, in the case where $\beta=0$, we have $\mathcal{C}=-4\pi r_h$ (the heat capacity of the BTZ black hole in Schwarzschild-type form) 
and for $0< r_h <r_{min}=\dfrac{1}{4}\sqrt{\dfrac{3}{2}}\beta$ the curves enter the stability region with $\mathcal{C}>0$.

\section{BTZ Gravastar with exponential distribution}
\label{stbtzgrav}
In this section, we will build a gravastar BTZ with the minimum length. For this, we consider an internal metric in which the static BTZ solution with the minimum length is contained; on its exterior, we have a BTZ metric, which is also static, both separated by a thin shell.
Hence, for the outer metric, we have ~\cite{Banados:1992wn}
\begin{equation}
ds^{2} = -f(r)_{+} dt_{+}^{2} + f(r)_{+}^{-1}dr_{+}^{2} + r_{+}^{2}d\phi_{+}^{2},
\end{equation}
where the metric function $f(r)_+$ in the outer region is
\begin{eqnarray}
\label{frl}
f(r)_+=-M_0 + \frac{r^{2}}{l^{2}}=M - \Lambda r^2,  
\end{eqnarray}
where $M=-M_0$ and $\Lambda=-1/l^2$ is the cosmological constant.

Now we express the metric function $f(r)_+$ in Schwarzschild-type form as follows:
\begin{eqnarray}
f(r)_{+} = 1- \frac{b_{+}}{r},
\end{eqnarray}
\begin{eqnarray}
b_+=-r\left(M - \Lambda r^2\right) +r.
\end{eqnarray}
                                 
For the inner region, the line element describing the three-dimensional spacetime is given by  
\begin{eqnarray}
ds^{2} =  - g(r)_{-}dt_{-}^{2} + f(r)_{-}^{-1}dr_{-}^{2} + r_{-}^{2}d\phi_{-}^{2},
\label{btzcm}
\end{eqnarray}
where the metric functions $g(r)_{-}$ and $f(r)_{-}$ are given respectively by
\begin{eqnarray}
g(r)_{-}&=& M + \left[\frac{(-8Mr+2\gamma)}{\gamma}\exp\left(\frac{-4r}{\gamma}\right)\right] -\Lambda r^2.
\label{a05}
\\
f(r)_{-}&=& 1- \frac{b_{-}}{r}, 
\end{eqnarray}      
being
\begin{eqnarray}
b_{-} = -r M  \left[1 - \frac{(4r + \gamma)}{\gamma} \exp\left(\frac{-4r}{\gamma}\right) \right]  + r,
\end{eqnarray}
and $\pm$ represents the outer and inner geometry, respectively. 
						
The distributions, both internally and externally, are enclosed by isometric hypersurfaces referred to as $\Sigma_{+}$ and $\Sigma_{-}$. Our objective is to connect $M_{+}$ and $M_{-}$ in their respective boundaries in order to attain a unified variety known as $\mathscr{M}$ such that $\mathscr{M} = M_{+} \cup M_{-}$, thereby ensuring that, in these boundaries, $ \Sigma = \Sigma_{+} = \Sigma_{-}$. As a result, to compute the elements of the energy-momentum tensor, we will employ the intrinsic metric in (2+1)-dimensions as follows~\cite{Lobo:2015lbc}:
\begin{equation}
ds^{2}_{\Sigma} = -d\tau^{2} + a(\tau)^{2} d\theta^{2}. 
\end{equation} 
Such that on the junction surface, $x^{\nu} (\tau,\theta,\phi) = (t(\tau), a(\tau), \theta)$ 
and the unit vectors to this surface are given by~\cite{MartinMoruno:2011rm}:			
\begin{equation}
n^{\mu}_{+} = \left(\frac{1}{M +\frac{a^{2}}{l^{2}}}\dot{a},\sqrt{M  +\frac{a^{2}}{l^{2}} + \dot{a}^{2}}, 0 \right), 
\end{equation}
\begin{equation}
n^{\mu}_{-} = \left(\frac{1}{M  - \left[\frac{M(4a + \gamma)}{\gamma}\exp\left(\frac{-4a}{\gamma}\right)\right]}\dot{a},\sqrt{M  - \left[\frac{M(4a +\gamma)}{\gamma}\exp\left(\frac{-4a}{\gamma}\right)\right]  + \dot{a}^{2}}, 0 \right), 
\end{equation}
where are using the definition $\dot{a}\equiv da/d\tau$. 

Next, to calculate the extrinsic curvatures we apply the equations given below~\cite{Visser:1995cc}: 
\begin{equation}
K^{\psi \pm}_{\,\ \psi} = \frac{1}{a}\sqrt{1 - \frac{b_{\pm}(a)}{a} + \dot{a}^{2}}, 
\end{equation}
and					
\begin{equation}
K^{\tau\pm}_{\,\ \tau} = \left\lbrace \frac{ \ddot{a} + \frac{ b_{\pm}(a) - b^{'}_{\pm}(a)a}{2a^{2}} }{\sqrt{1 - \frac{b_{\pm}(a)}{a} + \dot{a}^{2}}} + \Phi^{'}_{\pm}(a)\sqrt{1 - \frac{b_{\pm}(a)}{a} + \dot{a}^{2}}\right\rbrace .
\end{equation}
Therefore, using the Lanczos equations leads to the surface energy-momentum tensor in the following form~\cite{Lobo:2015lbc}:
\begin{equation}
S^{i}_{\, j} = - \frac{1}{8 \pi} (k^{i}_{\, j} - \delta^{i}_{\, j} \,\ k^{l}_{\,\, l}), 
\end{equation}
where in the expression above, $k^{i}_{\, j}=K^{i\, +}_{\, j} - K^{i\, -}_{\, j}$ is the 
discontinuity of the extrinsic curvature and the surface energy-momentum tensor can be written as 
$S^{i}_{\,j} = diag( - \sigma, \mathscr{P})$, where $\sigma$ is the surface density and $\mathscr{P}$ is the surface pressure~\cite{Perry:1991qq}, which are defined as  
\begin{equation}
\sigma = - \frac{K^{\psi }_{\,\ \psi}}{4 \pi} = - \dfrac{1}{4 \pi a} \left[ \sqrt{M -\Lambda a^{2} + \dot{a}^{2}} - \sqrt{M  - \left[\frac{M(4a + \gamma)}{\gamma}\exp\left(\frac{-4a}{\gamma}\right)\right] + \dot{a}^{2}} \,\ \right], 
\label{a1}
\end{equation}							
\begin{equation}
 \mathscr{P} =  \frac{K^{\tau }_{\,\ \tau} + K^{\psi }_{\,\ \psi} }{8 \pi } 
= \frac{1}{8  \pi a} \left\{ \frac{M + \dot{a}^{2} +  \ddot{a} - 2 \Lambda a^{2}}{\sqrt{M - \Lambda a^{2} + \dot{a}^{2}}} - \frac{M  + \dot{a}^{2} +  \ddot{a} + M \exp\left(\frac{-4 a}{\gamma}\right)  \left[  \dfrac{4 a^{2}}{\gamma^{2}} - \dfrac{4 a}{\gamma} - 1\right]}{\sqrt{M  - \left[\frac{M(4a + \gamma)}{\gamma}\exp\left(\frac{-4a}{\gamma}\right)\right]  + \dot{a}^{2}} } \right\}.
\label{a2}
\end{equation}
Furthermore, considering a static solution, the equations for $\sigma$ and $\mathscr{P}$ become
\begin{equation}
\sigma (a_0)  = - \frac{K^{\psi }_{\,\ \psi}}{4 \pi} = - \dfrac{1}{4 \pi a_{0}} \left[ \sqrt{M - \Lambda a^{2}_{0}} - \sqrt{M  - \frac{M(4a_{0}+ \gamma)}{\gamma}\exp\left(\frac{-4a_{0}}{\gamma}\right)} \,\ \right], 
\label{a3}
\end{equation}							
\begin{equation}
\mathscr{P} (a_0)  =  \frac{K^{\tau }_{\,\ \tau} + K^{\psi }_{\,\ \psi} }{8 \pi } 
=\frac{1}{8  \pi a_{0}} \left[ \frac{M  - 2 \Lambda a_{0}^{2}}{\sqrt{M - \Lambda a_{0}^{2}}} - \frac{M + M \exp\left(\frac{-4 a_{0}}{\gamma}\right) \left[ \dfrac{4 a_{0}^{2}}{\gamma^{2}} - \dfrac{4 a_{0}}{\gamma} -1 \right]}{\sqrt{M  - \left[\frac{M(4 a_{0}+\gamma)}{\gamma}\exp\left(\frac{-4a_{0}}{\gamma}\right)\right]}} \right]. 
\label{a4}
\end{equation}
At this point, we introduce the dimensionless quantities $\Tilde{\Lambda} = \Lambda a^2_0$ and $\eta ={\gamma}/{a_{0}}$, such that the above equations take the following form:
 \begin{equation}
\Tilde{\sigma}  = - \dfrac{1}{4 \pi} \left[ \sqrt{M - \Tilde{\Lambda} } - \sqrt{M  - \left[M\left(\dfrac{4}{\eta}+1\right)\exp\left(\dfrac{-4}{\eta}\right)\right]} \,\ \right],
\label{a5}
\end{equation}							
\begin{equation}
\Tilde{\mathscr{P}} =\frac{1}{8 \pi} \left\{ \frac{M - 2 \Tilde{\Lambda}}{\sqrt{M - \Tilde{\Lambda}}} - \frac{M + M \exp\left(\dfrac{-4}{\eta}\right) \left( \dfrac{4}{\eta^{2}} - \dfrac{4}{\eta} + 1 \right)}
{\sqrt{M - M\left( \dfrac{4}{\eta} - 1\right) \exp\left(\dfrac{-4}{\eta}\right)}} \right\},
\label{a6}
\end{equation}                                    
where we have defined $\Tilde{\sigma}= a_0\sigma(a_0) $ and $\Tilde{\mathscr{P}}= a_0\mathscr{P}(a_0)$. 
\begin{figure}[tb]	
\centering 
\includegraphics[scale=.26]{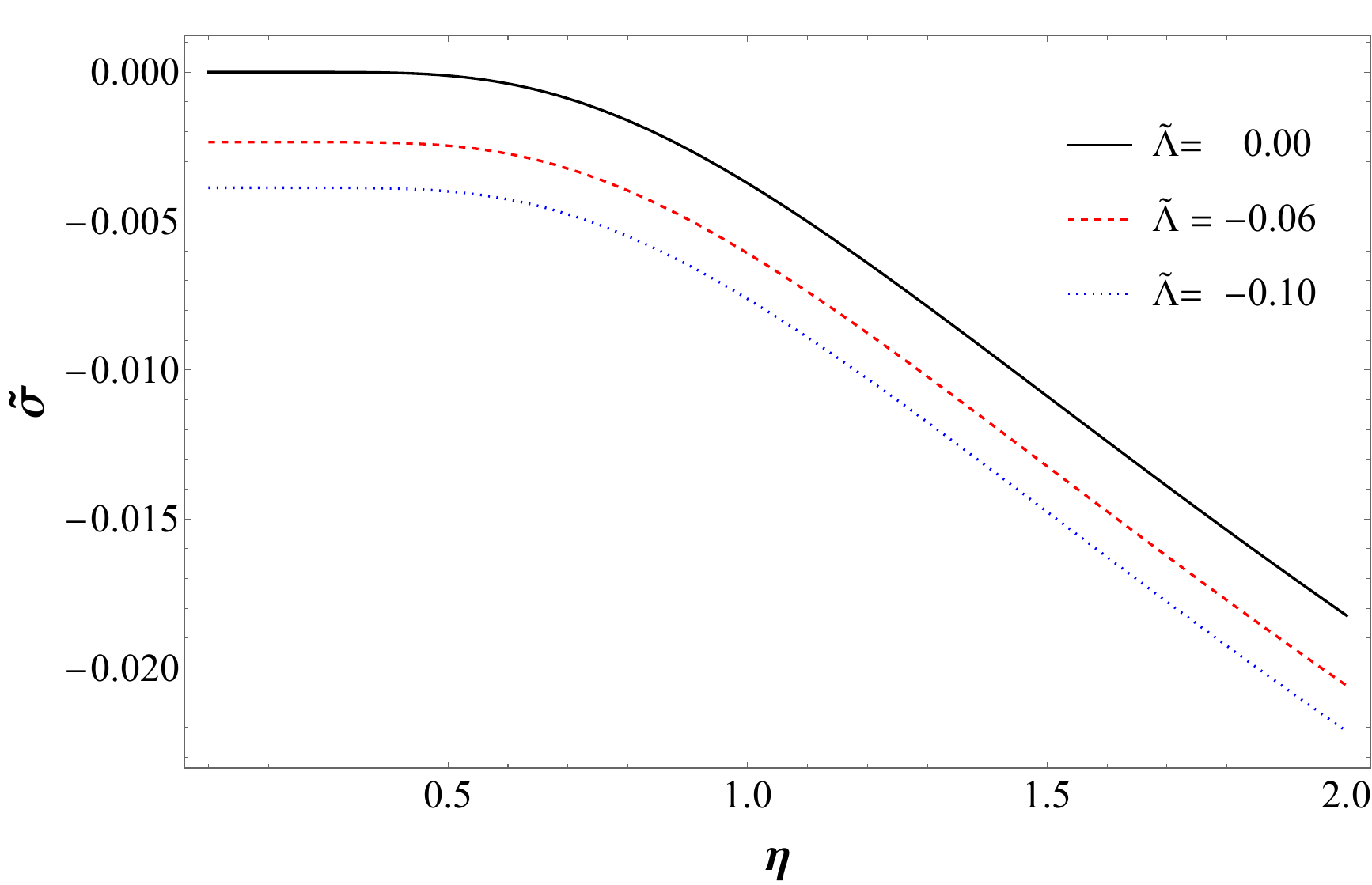} 
\caption{Energy density $\Tilde{\sigma}$ as a function of parameter $\eta$ for $\Tilde{\Lambda} = 0$ (black, solid), $\Tilde{\Lambda} = -0.06$ (red, dashed) and $\Tilde{\Lambda} = -0.10$ (blue, dotted), assuming $ M=1 $.}
\label{G1a}
\end{figure}  
\begin{figure}[h]	
\centering 
\includegraphics[scale=.25]{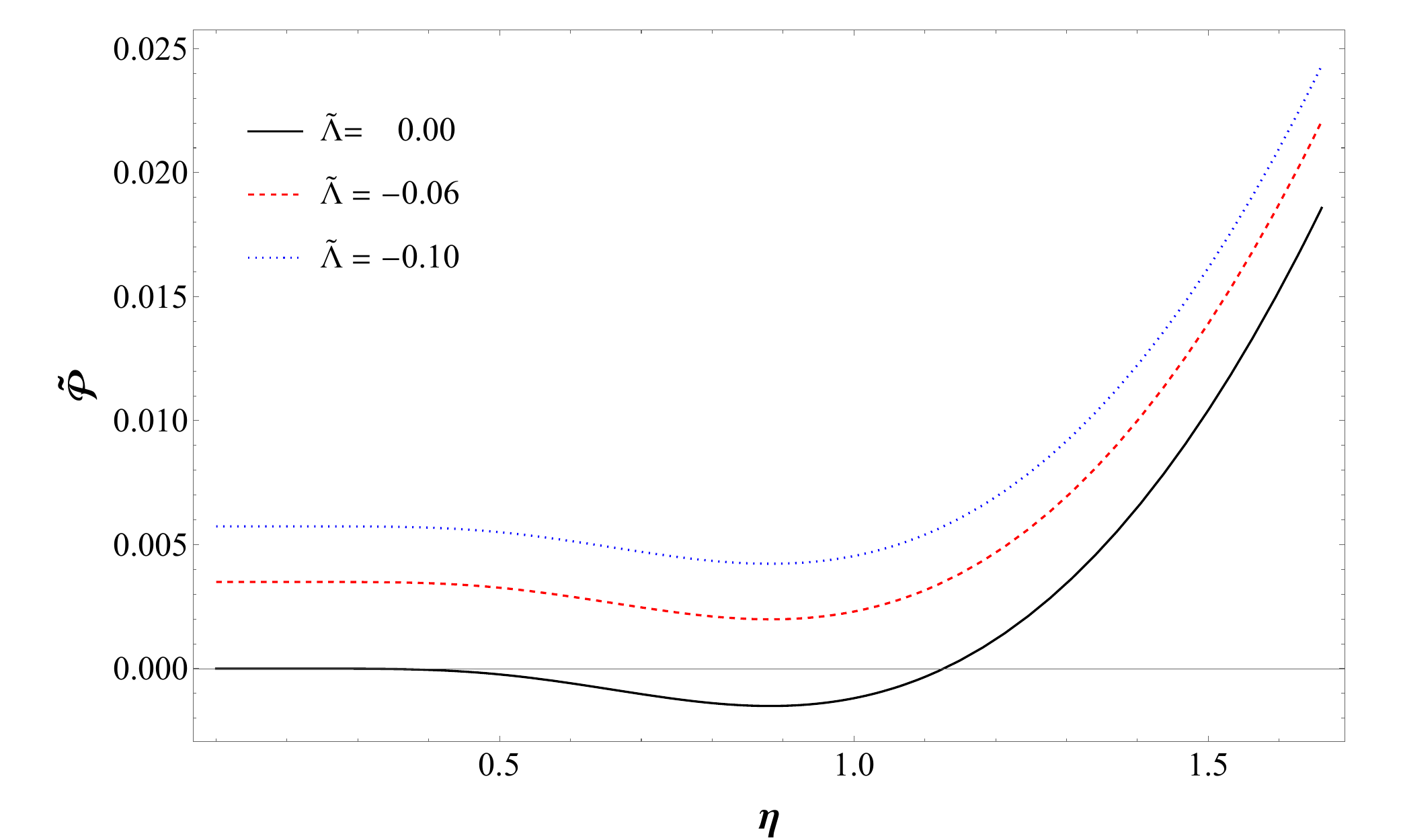} 
\caption{Pressure $\tilde{\mathscr{P}}$ as a function of parameter $\eta$ for  $\Tilde{\Lambda} = 0$ (black, solid), $\Tilde{\Lambda} = -0.06$ (red, dashed) and $\Tilde{\Lambda} = -0.10$ (blue, dotted), assuming $ M=1 $.}
\label{G2a}
\end{figure} 

Now, for $\Lambda=0$ and $\eta\, (\gamma \ll 1)\rightarrow 0$, we have $\exp (-4 /\eta)=b \ll 1$. So, we can rewrite the equations \eqref{a5} and \eqref{a6}: 
 \begin{eqnarray}
\Tilde{\sigma} =-\frac{\sqrt{M}}{4\pi} \left[1 - \sqrt{1 - \left(\frac{4}{\eta} + 1 \right) b}\right] 
\approx -\frac{\sqrt{M} b}{2\pi\eta} = - \frac{\sqrt{M}}{2\pi\eta}e^{-4 /\eta}.
\end{eqnarray}
Then
\begin{equation}
{\sigma} \approx - \frac{\sqrt{M}}{2\pi a_0\eta}e^{-4 /\eta},
\end{equation}
and
\begin{equation}
\tilde{\mathscr{P}}= \frac{\sqrt{M}}{{8 \pi}} \left[1 - \frac{1}{\sqrt{1- \left(\frac{4}{\eta}-1\right)b}} -\frac{b\left( \frac{4}{\eta^2} -\frac{4}{\eta} + 1\right)}{\sqrt{1- \left(\frac{4}{\eta}-1\right)b}} \right] \approx -\frac{\sqrt{M}}{{2\pi\eta^2}}\, e^{-4/\eta} .
\end{equation}  
Hence
\begin{eqnarray}
 {\mathscr{P}}\approx  -\frac{\sqrt{M}}{{2\pi a_0 \eta^2}}\, e^{-4/\eta}
 =-\frac{a_0\sqrt{M}}{{2\pi\gamma^2}}\, e^{-4 a_0/\gamma} . 
 \label{pshell}
\end{eqnarray}
Therefore, in this approximation, we have found the following relationship: ${\mathscr{P}}={\sigma}/{\eta}$ in the thin shell. 
In Fig.~\ref{G1a}, we show the behavior of the energy density as a function of the parameter $\eta$. Thus, we observe that the energy density is negative both for $\tilde{\Lambda}<0$ as well as for $\tilde{\Lambda}=0$.
In Fig.~\ref{G2a}, we show the behavior of pressure as a function of the parameter $\eta$. Hence, as observed in the graph, the pressure is positive for $\tilde{\Lambda}<0$.
However, when considering $\tilde{\Lambda} = 0$, there is a region where $\mathscr{P} < 0$. This implies that for this specific case, the equation of state ${\mathscr{P}}=-{\sigma}=\rho$ is not satisfied, as illustrated in equation (\ref{pshell}) and in Fig. \ref{G2a}. 
Furthermore, for $\tilde{\Lambda} = 0$, the minimum length parameter $\gamma$ does not play the role of the cosmological constant for gravastar formation and stability.

In mechanical systems, the disorder is quantified by entropy. In the thin shell, the region bounded between $r_1=a_0$ and $r_2=a_0 + \epsilon$ with $0<\epsilon\ll 1$, according to the Mazur-Mottola model \cite{Mazur:2001fv, Mazur:2004fk}. Based on the entropy function of the following form, it is assessed~\cite{Bhattacharjee:2025wva, Bhattacharjee:2024xjf}:
\begin{equation}
S_{shell}=2\pi\int_{r_{1}}^{r_{2}} \mathfrak{s}(r)r \dfrac{dr}{\sqrt{f(r)}} . \label{eq44}
\end{equation}
Here $\mathfrak{s}(r)=\frac{\omega^2k_{B}^2T(r)}{4\pi\hbar^2}=\frac{\omega k_B}{\hbar}\sqrt{\frac{p}{2\pi}}$ is the entropy density corresponding to a local specific temperature $T(r)$, $\omega$ is a dimensionless constant, and we consider $\omega=1$ without any loss of generality, $k_{B}$ is the Boltzmann constant and $\hbar=\frac{h}{2\pi}$ is the Planck constant. 
Then, by using (\ref{a05}) and (\ref{pshell}) in \eqref{eq44}, we obtain the total entropy in the Planckian units $(\hbar=k_{B}=1)$ as: 
\begin{equation}
S_{shell}=2\pi\int_{r_{1}}^{r_{2}} \sqrt{-\dfrac{r\sqrt{M}\, e^{-4 r/\gamma}}{4 \pi^2 \gamma^2}} \dfrac{r}{\sqrt{ M  \left[1 - \frac{(4r + \gamma)}{\gamma} \exp\left(\frac{-4r}{\gamma}\right) \right] } }~dr. 
\label{eq45}
\end{equation}
Then, for $\gamma\ll 1$, we have
\begin{equation}
S_{shell}\approx\sqrt{\frac{1}{4 \gamma\sqrt{M}}}\int_{r_{1}}^{r_{2}} r ~dr
\approx \sqrt{\frac{1}{4 \gamma\sqrt{M}}}\, \left[\frac{r^2}{2}\right]_{r_1=a_0}^{r_2=a_0 + \epsilon}
\approx \frac{\epsilon}{2}\sqrt{\frac{1}{4 \gamma\sqrt{M}}}(2a_0 + \epsilon). 
\label{eq45a}
\end{equation}
\begin{figure}[h!]	
\centering 
\includegraphics[scale=.35]{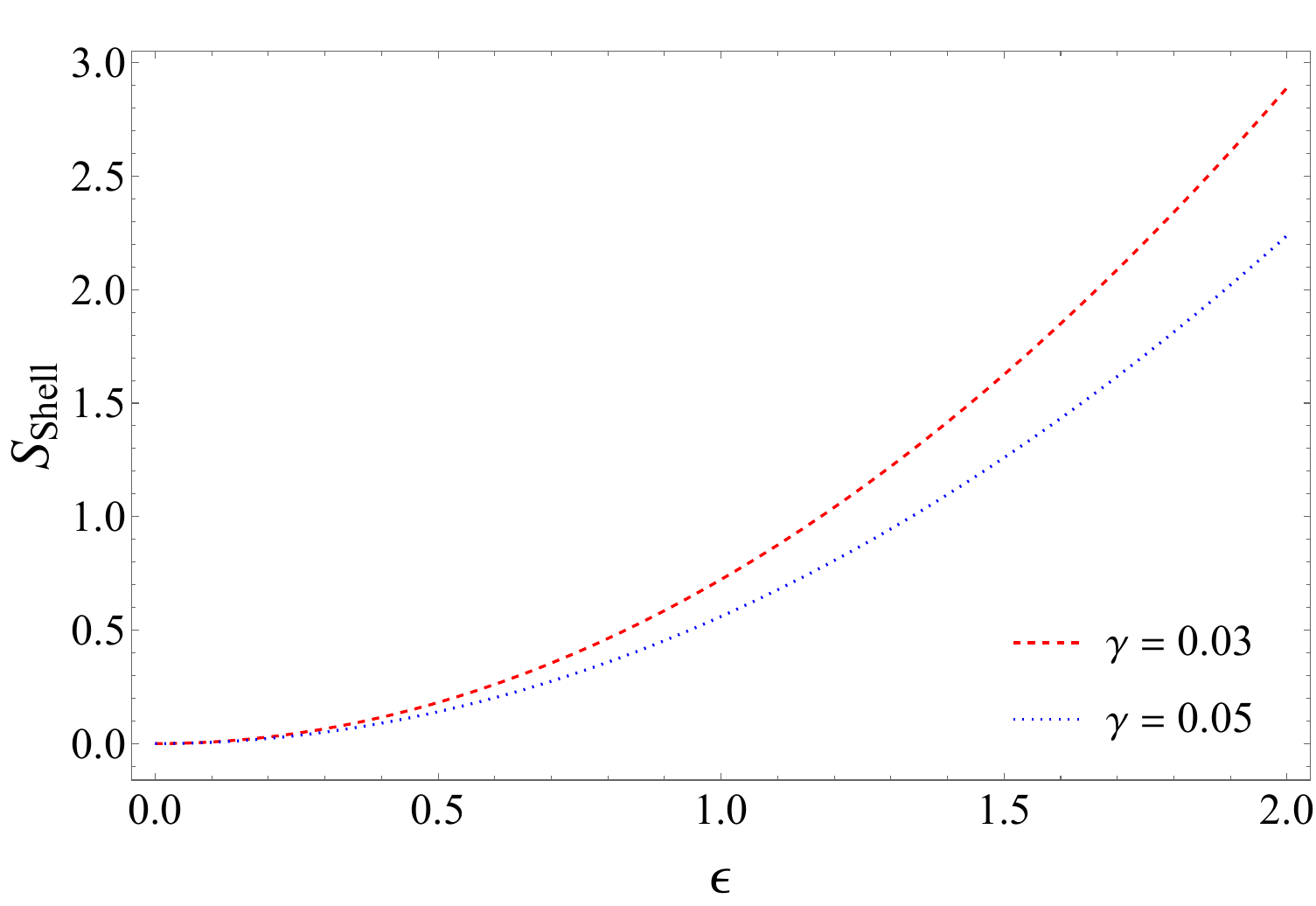} 
\caption{Variation of the entropy ($S_{shell}$) with respect to Shell thickness ($\epsilon$) for $M = 1$, $a_{0} = 0.001$, $\gamma = 0.03$ (red, dashed) and $\gamma = 0.05$ (blue, dotted)}
\label{entro1}
\end{figure} 

When examining Eq. (\ref{eq45}) and Fig. \ref{entro1}, the fundamental importance of the minimum length parameter, $\gamma$, in determining the entropy of the gravastar shell is highlighted. It is noted that, as  $\gamma \rightarrow 0$, the entropy diverges, demonstrating the physical impossibility of this limit. 
This divergence emphasizes the need for a correction parameter to maintain the structural stability of the gravastar in the absence of the cosmological constant. 
This is connected with the result presented in Fig.~\ref{G2a} for the case $\Lambda=0$, indicating that in the region where $\gamma$ is small, the condition $\mathscr{P}=\rho$ is not satisfied and thus the parameter $\gamma$ does not play the role of the cosmological constant for the formation and stability of a gravastar.

\section{BTZ Gravastar with Lorentzian-Type Distribution}
\label{LTDist}
In this section, we follow the same steps described in the previous section, but now considering a line element in the interior region given by~\cite{Anacleto:2022sim}
\begin{eqnarray}
ds^{2} =  -g(r)_{-}dt_{-}^{2} + f(r)_{-}^{-1}dr_{-}^{2} + r_{-}^{2}d\phi_{-}^{2}.
\label{metcs}
\end{eqnarray}
Here the metric functions $g(r)_{-}$ and $f(r)_{-}$ are written respectively as
\begin{eqnarray}
g(r)_{-} &=& - \frac{16Mr^2}{(\beta + 4 r)^2} + \dfrac{r^{2}}{l^{2}},
\label{fmetcs}
\\
f(r)_{-} &=& 1- \frac{b_{-}}{r}, 
\end{eqnarray}
where
\begin{eqnarray}
b_{-}=-r\, \left(\frac{16Mr^2}{(\beta + 4 r)^2}\right) + r.   
\end{eqnarray}
Now, using the Lanczos equation and doing some algebraic manipulations, we can obtain the energy density and the surface pressures which are given by
 \begin{equation}
\sigma = - \frac{K^{\psi }_{\,\ \psi}}{4 \pi} = - \dfrac{1}{4 \pi a} \left[ \sqrt{M - \Lambda a^{2} + \dot{a}^{2}} - \sqrt{\frac{16M a^2}{(\beta + 4 a)^2} + \dot{a}^{2}} \,\ \right], \label{t1}
\end{equation}							
\begin{equation}
\mathscr{P} =  \frac{K^{\tau }_{\,\ \tau} + K^{\psi }_{\,\ \psi} }{8 \pi } 
 =\frac{1}{8 \pi a} \left[\frac{M +\dot{a}^{2} +  \ddot{a} - 2\Lambda a^{2}}{\sqrt{M - \Lambda a^{2} + \dot{a}^{2}}} -\dfrac{ \dfrac{32 M a^2}{(4 a + \beta)^2}  + \dot{a}^{2} + \ddot{a} - \dfrac{64 M a^3}{(4 a + \beta)^3} }{\sqrt{\dfrac{16M a^2}{(\beta + 4 a)^2} + \dot{a}^{2}} }  \right]. 
\label{t2}
\end{equation}
Considering the static case, $a_{0} \in (r_{-},r_{+})$, for a better discussion, we have:
\begin{equation}
\sigma(a_{0})  = - \dfrac{1}{4 \pi a_{0}} \left[ \sqrt{M - \Lambda a^{2}_{0}} - \sqrt{\frac{16M a_{0}^2}{(\beta + 4 a_{0})^2}} \,\ \right], 
\label{t3}
\end{equation}							
\begin{equation}
\mathscr{P}({a_0}) =\frac{1}{8 \pi a_{0}} \left[\frac{M - 2\Lambda a^{2}_{0}}{\sqrt{M - \Lambda a^{2}_{0} }} -\dfrac{ \dfrac{32 M a_{0}^2}{(4 a_{0} + \beta)^2}  - \dfrac{64 M a_{0}^3}{(4 a_{0} + \beta)^3} }{\sqrt{\dfrac{16M a_{0}^2}{(\beta + 4 a_{0})^2} } }  \right]. 
\label{silva1}
\end{equation}    
We can rewrite the equation (\ref{silva1}):
\begin{equation}
\mathscr{P}({a_0})\approx\frac{1}{8 \pi a_{0}} \left[\frac{M - 2\Lambda a^{2}_{0}}{\sqrt{M - \Lambda a^{2}_{0} }} - \dfrac{1}{\sqrt{M}} \left(\dfrac{8 M a_{0}}{(4 a_{0} + \beta)}  -  M \right)  \right]. 
\end{equation}
Now we can write the above equations in terms of the dimensionless parameters $\Tilde{\Lambda} = \Lambda a^2_0$ 
and $\alpha={\beta}/{a_{0}}$ as follows:   
\begin{equation}
\Tilde{\sigma}= - \dfrac{1}{4 \pi} \left[ \sqrt{M - \Tilde{\Lambda}} - \sqrt{\frac{M}{(1 + \frac{\alpha}{4})^2}} \,\ \right], 
\label{t3}
\end{equation}							
\begin{equation}
\Tilde{\mathscr{P}} =\frac{1}{8 \pi } \left[\frac{M - 2\Tilde{\Lambda}}{\sqrt{M - \Tilde{\Lambda}}} - \dfrac{1}{\sqrt{M}}\left( \dfrac{2 M}{(1 +\frac{\alpha}{4})}- M\right) \right]. 
\label{t4}
\end{equation}
\begin{figure}[!h]	
\centering 
\includegraphics[scale=.35]{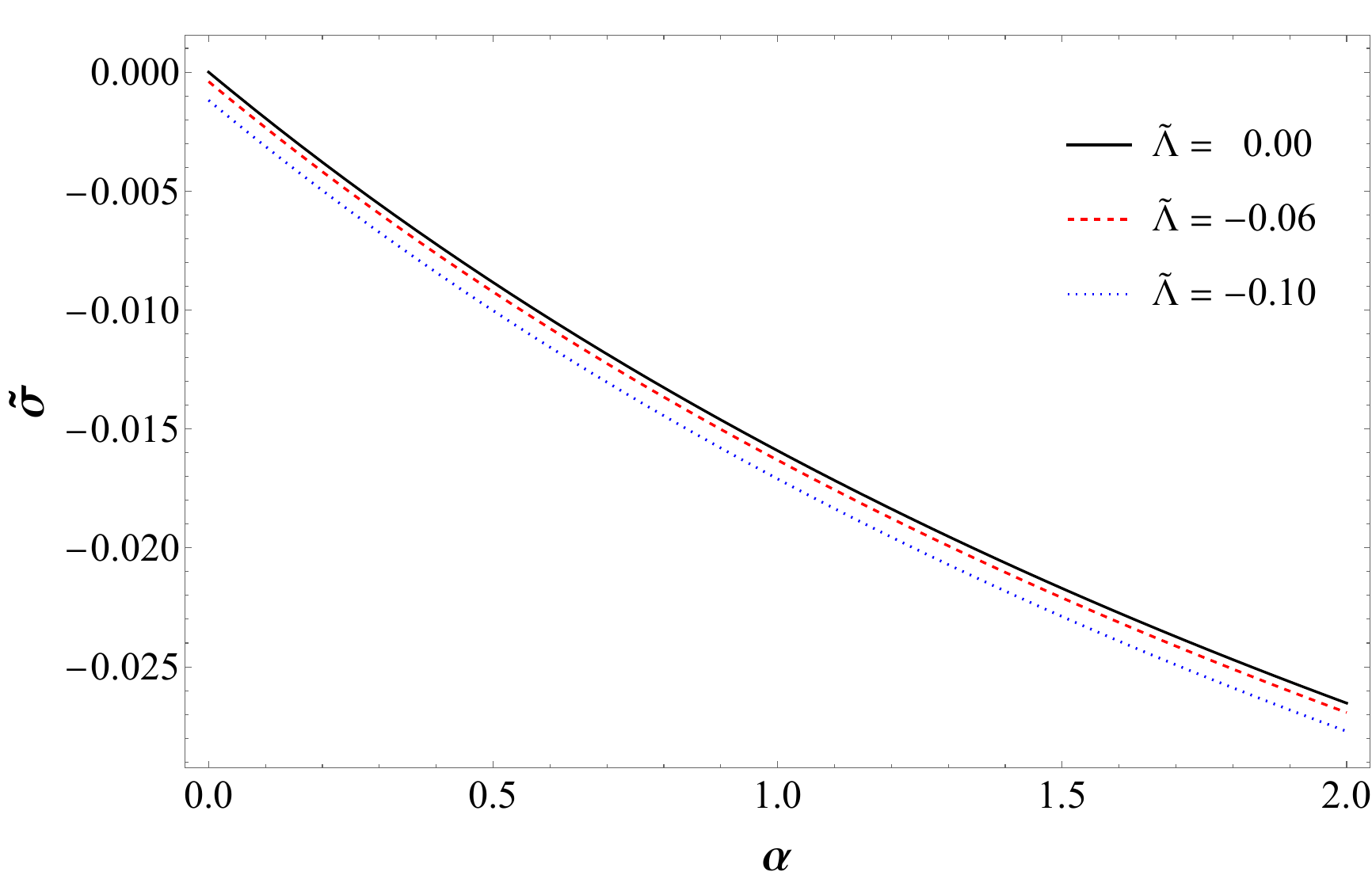} 
\caption{Energy density $\Tilde{\sigma}$ as a function of parameter $\alpha$ for $\Tilde{\Lambda} = 0$ (black, solid), $\Tilde{\Lambda} = -0.06$ (red, dashed) and $\Tilde{\Lambda} = -0.10$ (blue, dotted), assuming $ M=1 $.}
\label{G3a}
\end{figure}     
\begin{figure}[h!]	
\centering 
\includegraphics[scale=.38]{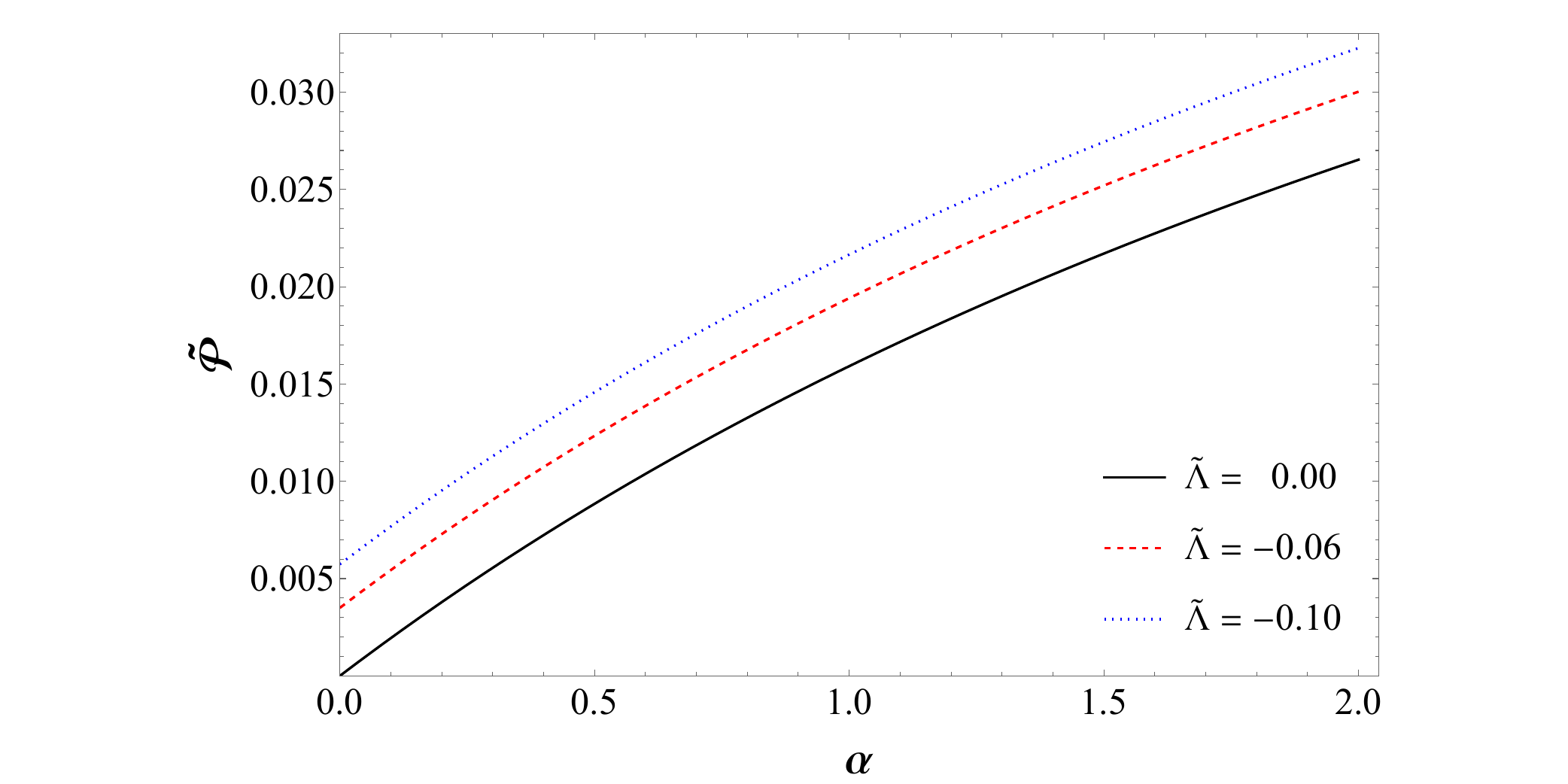} 
\caption{Pressure $\tilde{\mathscr{P}}$ as a function of parameter $\alpha$ for  $\Tilde{\Lambda} = 0$ (black, solid), $\Tilde{\Lambda} = -0.06$ (red, dashed) and $\Tilde{\Lambda} = -0.10$ (blue, dotted), assuming $ M=1 $.}
\label{G4a}
\end{figure} 

At this point, we will analyze the case where $\Lambda=0$, and so the above equations for energy density and pressure become
\begin{eqnarray}
&&\Tilde{\sigma}= - \dfrac{1}{4 \pi} \left[ \sqrt{M } -  \sqrt{\frac{M}{(1 + \frac{\alpha}{4})^2}} \,\ \right], 
\label{t5}
\\
&&\Tilde{\mathscr{P}} =\frac{1}{8 \pi } \left[\frac{M}{\sqrt{M}} - \dfrac{1}{\sqrt{M}} \left(\dfrac{2M}{(1 + \frac{\alpha}{4}) } - M\right)\right]. 
\label{t6}
\end{eqnarray}
Therefore, analyzing the result above, we verify that the conditions $\Tilde{\sigma}<0$ and $\Tilde{\mathscr{P}}>0$ are maintained due to the presence of the minimum length.
Now, for $\alpha\ll 1\, (\beta\ll 1)$ we find
\begin{eqnarray}
&&{\sigma}\approx - \dfrac{\sqrt{M}\alpha}{16 \pi a_0} = - \dfrac{\sqrt{M} \beta}{16 \pi a_{0}^{2}}, 
\label{a10}
\\
&&{\mathscr{P}}\approx \dfrac{\sqrt{M}\alpha}{16 \pi a_0}=\dfrac{\sqrt{M} \beta}{16 \pi a_{0}^{2}}.
\label{a11}  
\end{eqnarray}   
Moreover, we obtain the equation of state in the thin shell
\begin{eqnarray}
{\sigma} + {\mathscr{P}} =0,\, \quad\quad 
{\mathscr{P}}=-{\sigma}=\rho.
\end{eqnarray}
Note that the equation of state, $\mathscr{P}=\rho$, arises only when $\beta\neq 0$. 
However, for $\Lambda=0$, inside the shell we have $\mathscr{P}=-\rho$ with $\rho\approx\beta/M_0 r^3$ acting as a repulsive pressure. 
Furthermore, this would be related to dark energy arising due to the effect of the minimum length
(\textit{$\beta$-dark energy}).  
Here, we show that by setting $\Lambda=0$, the minimum length parameter $\beta$ plays the role of the cosmological constant for gravastar formation and stability. 
This behavior can be directly observed in Figs.~\ref{G3a} and~\ref{G4a} for $\tilde{sigma}<0$ and $\mathscr{P}>0$.

On the other hand, by considering $\Tilde{\Lambda}$ too large, we can write the equations for $\Tilde{\sigma}$ and $\Tilde{\mathscr{P}}$ as follows:
\begin{eqnarray}
&&{\sigma} \approx -\frac{\sqrt{-\Tilde{\Lambda}}}{4 \pi a_0}=-\frac{\sqrt{{-\Lambda a^2_0}}}{4 \pi a_0}, 
\label{alambg}
\\
&&{\mathscr{P}} \approx \frac{\sqrt{-\Tilde{\Lambda}}}{4 \pi a_0}=\frac{\sqrt{{-\Lambda a^2_0}}}{4 \pi a_0}.
\label{plambg}  
\end{eqnarray}
For this case, with ${\Lambda}<0$, we obtain the following equation of state
\begin{eqnarray}
{\mathscr{P}}=-{\sigma}.
\end{eqnarray}
 By comparing the results above, we find a relationship between ${\Lambda}$ and $\beta$ given by 
\begin{eqnarray}
\sqrt{-{\Lambda}a^2_0} = \dfrac{\sqrt{M}\alpha}{4}=\dfrac{\sqrt{-M_0}\beta}{4a_{0}}, \qquad \mbox{or} \qquad 
\beta = \sqrt{\dfrac{16  a^{4}_{0} \Lambda}{M_{0}}}.
\end{eqnarray}
In order to obtain a value for the $\beta$ parameter, we consider $M_{0} = {M_{BH}}/{M_{\odot}}$, with
$M_{BH} $ being the mass of the black hole and $M_{\odot}=1.989 \times 10^{30}$ kg the solar mass. 
So for $M_{BH}=10M_{\odot}$, $a_0\approx 29.5 \times 10^{3}$ m (radius of the black hole) and 
$\Lambda=1.088 \times 10 ^{-58}$ m$^{-2}$ (cosmological constant), we find 
\begin{eqnarray}
 \beta \approx 1.15 \times 10^{-20} m =\left[0.583 \times 10^{4} ~ GeV\right]^{-2}
=\left[0.583 \times 10\, TeV \right]^{-2} .
\end{eqnarray}
Hence, we have obtained a value for the parameter $\beta\sim [10$ TeV$]^{-2}$ or $ \sqrt{\beta}\sim [10$ TeV$]^{-1}$,
with an energy scale $\Lambda_{ml}=1/\sqrt{\beta}\sim 10$ TeV. 
The result obtained is in agreement with those found in the literature~\cite{Mocioiu:2000ip,Falomir:2002ih,Vagnozzi:2022moj} 
and also in the context of the thin-shell gravastar model in a noncommutative BTZ geometry~\cite{silva2024}.

Now, we can find the Entropy within the shell, by~\cite{Bhattacharjee:2025wva, Bhattacharjee:2024xjf}
 \begin{equation}
S_{shell}=2\pi\int_{r_{1}}^{r_{2}} \mathfrak{s}(r)r \dfrac{dr}{\sqrt{f(r)}} . \label{eq49}
\end{equation}
By using (\ref{a05}) and (\ref{a11}) in \eqref{eq49}, we obtain the total entropy in the Planckian units $(\hbar=k_{B}=1)$ as: 
\begin{eqnarray}
S_{shell}&=&\frac{1}{8}\sqrt{\frac{\sqrt{M}}{M}}\int_{r_{1}}^{r_{2}} (\beta + 4 r)\sqrt{\frac{1- 2\Lambda r^2/M}{r\sqrt{1- \Lambda r^2/M}} - \frac{1}{r} + \frac{\beta}{2 r^2}}~dr,
\\
&\approx & \frac{1}{8}\sqrt{\frac{\sqrt{M}}{M}}\int_{r_{1}}^{r_{2}}
\left[\frac{4\sqrt{\beta}}{\sqrt{2}} - \frac{\Lambda\sqrt{\beta} r^2}{M\sqrt{2}} -  \frac{4 \Lambda r^3}{M\sqrt{2\beta}}\right] ~dr,
\\
&=&\frac{\epsilon\sqrt{\beta M}}{2M\sqrt{2}} - \frac{\epsilon\Lambda}{M}\left(\frac{a^2_0\sqrt{M\beta}}{8M\sqrt{2}}
+ \frac{a^3_0\sqrt{M}}{2M\sqrt{2\beta}} \right) + \mathcal{O}(\epsilon^2).
\end{eqnarray}
\begin{figure}[h!]	
\centering 
\includegraphics[scale=.3]{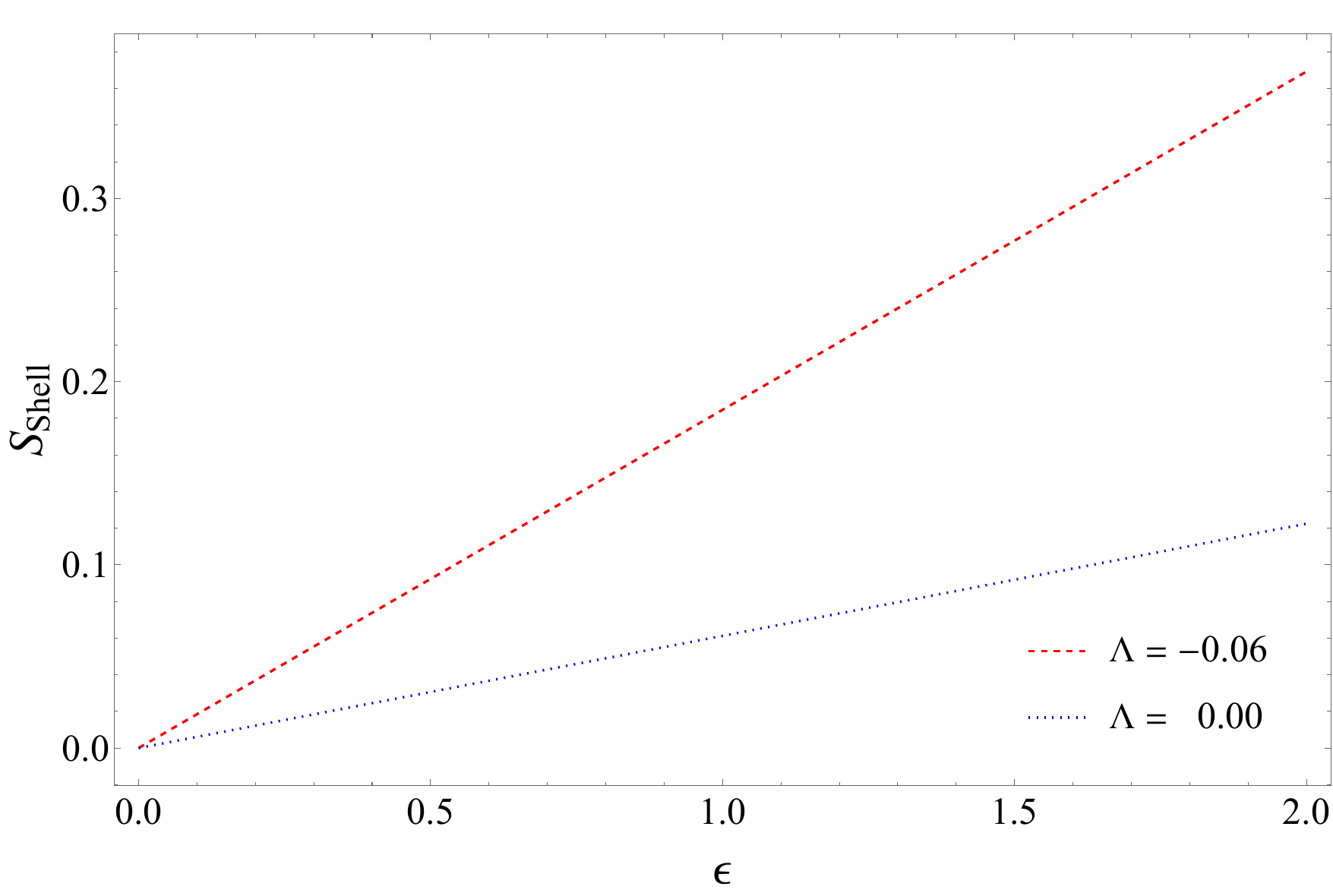} 
\caption{Entropy variation ($S_{shell}$) as a function of the shell thickness ($\epsilon$) for $M=1$, $a_{0}=1$, and $\beta=0.03$, considering $\Lambda=-0.06$ (red dashed) and $\Lambda=0$ (blue dotted).}
\label{entroLo1}
\end{figure} 
The equation above clearly demonstrates the mathematical relationship that exists between the entropy of the shell $S_{shell}$ and the thickness $\epsilon$. 
In Eq. (\ref{eq49}) and Fig.~\ref{entroLo1}, we can see that the cosmological constant $\Lambda$ plays a fundamental role in entropy. However, in the case where we have $\Lambda = 0$, the parameter $\beta$ becomes essential to characterize the thermodynamic properties and stability of the shell. This suggests that variations in these parameters can significantly impact the behavior of the model.

It is important to highlight that the current construction is sufficiently comprehensive to accommodate a recording configuration composed of three distinct layers, each with its own equations of state. In this scenario, the introduction of a minimum length alters the underlying geometry, rather than establishing a specific thermodynamic relationship between pressure and energy density. As a consequence, the inner region can still be characterized by a modified anti-de Sitter type equation of state, while the outer region corresponds to a BTZ type spacetime in vacuum. Furthermore, the thin layer arises naturally from the Israel junction conditions, being compatible with (2+1)-dimensional geometries. Specifically, the layer's equation of state is not determined a priori; it arises from the geometry altered by the surface energy density and pressure. Thus, the existence of a minimum length does not prevent the occurrence of three regions with different equations of state; on the contrary, it maintains the structural requirements of a gravastar, allowing the inner, outer, and outer layers to have independent effective equations of state.

\section{DISCUSSION AND CONCLUSION} \label{conc}
In this paper, based on requirements related to small-scale physics, thermodynamic stability, and phenomenological energy scales, we restrict the ranges of the minimum length parameters. These parameters are represented by $\gamma$ in the exponential model and by $\beta$ in the Lorentzian model. In the exponential model, based on the probability density of the ground state of the hydrogen atom in two dimensions, small values are used for the parameter $\gamma$, such as $\gamma = 0.03$ and $\gamma = 0.05$, when choosing units where $l = 1$. This is because the presence of a minimum length is related to small-magnitude quantum corrections. In the limit where $\gamma \ll 1$, it is noted that the entropy of the gravastar shell remains finite. On the other hand, in the limit where $\gamma \to 0$, there is a divergence, which demonstrates the physical infeasibility of this regime and emphasizes the importance of maintaining $\gamma$ as a positive regulator. Even so, it is observed that this parameter, despite stabilizing the internal temperature, is not adequate to ensure the stability of the shell in small-scale regions when the cosmological constant is zero, since the equation of state $P = -\sigma$ is not satisfied in this situation.

In contrast, in the model based on the Lorentzian distribution, the parameter $\beta$ is introduced, which, in the limit  $\beta \ll 1$, plays the role of an effective repulsive pressure. This allows the gravastar to maintain its stability even without the presence of a cosmological constant. By comparing its effects with those related to the observational cosmological constant and taking into account a typical mass of  $10 M_{\odot}$, you estimate the value $ \beta \approx 1{,}15 \times 10^{-20} \, \text{m}$. This value is consistent with the phenomenological limits previously established in the literature, corresponding to a minimum length energy scale of $\Lambda_{ml} = 1/\sqrt{\beta} \approx 10 \, \text{TeV}$. Thus, while $\gamma$ functions as an essential regulator primarily to prevent thermodynamic divergences,  $\beta$ also plays a structural role in the stability of the configuration. This makes the Lorentzian model more robust in representing stable gravastars without the need for an external cosmological constant.

It is important to highlight that the matter content of the thin shell does not need to behave like a perfectly isotropic fluid, and in fact, that is not what is assumed here. Within the junction formalism, the shell is described by a surface energy density and a tangential pressure, but no radial pressure is defined along the hypersurface. This characteristic, far from being a limitation, naturally reflects the geometry of the problem and leads to an effectively anisotropic description of the shell's matter content. Below, we highlight the prominent characteristics of the gravastars analyzed in this study:
\begin{itemize}
\item \textbf{Interior Region:}
Initially, we examined the internal structures, emphasizing that the minimum length plays a regulatory role in the temperature of the Schwarzschild anti-de Sitter black hole in three dimensions, as observed in Figs.~\ref{fg0} and \ref{figth}. 
By calculating the entropy in the inner region, we identified correction terms present in Eqs. (\ref{entrexp}) and (\ref{entrlo}). Through the thermal capacity, we verified the stability of the black hole and discovered that, in both cases, there exists a minimum radius where $C \rightarrow 0$, indicating the formation of a black hole remnant as the final stage, as shown in Figs.~\ref{fg001} and \ref{fg002}. 
Comparing our results with Ref.~\cite{silva2024}, which addresses the solution of a gravastar in a noncommutative geometry, we observed significant similarities.
\item \textbf{Junction Condition:}  
For the formation of a thin shell, we applied the junction conditions between the internal and external spacetime, considering the minimum length distributions in both cases. 
Subsequently, we studied the behavior of the surface energy density $\sigma$ for different values of the cosmological constant $\Lambda$. 
When considering $\Lambda = 0$, we observed that the minimum length parameters take on the role of the cosmological constant, as illustrated in Figs.~\ref{G1a} and \ref{G3a}. 
Furthermore, we analyzed the behavior of the pressure, shown in Figs.~\ref{G2a} and \ref{G4a}, and identified the following patterns for each distribution:  
\begin{itemize}  
\item \textbf{Exponential Distribution:}  
 When considering $\Tilde{\Lambda} = 0$, we observed a small region where the pressure assumes negative values, as illustrated in Fig.~\ref{G2a}. This behavior can be interpreted as a local instability. By performing some approximations, 
 we arrived at Eq.~(\ref{pshell}) and identified the following relationship between energy density and pressure: 
 ${\mathscr{P}}={\sigma}/{\eta}$. However, this distribution does not satisfy the equation of state (EOS) 
${\mathscr{P}}=-{\sigma}=\rho$. {Although the system is thermodynamically stable in its interior, the minimum length parameter for this type of distribution is not sufficient to make it stable in its shell in small-scale regions.}
\item \textbf{Lorentzian Distribution:}  
 Unlike the previous distribution, in this case, we did not identify any local instability associated with the pressure when considering $\Tilde{\Lambda} = 0$, as properly illustrated in Fig.~\ref{G4a}. Moreover, after the necessary approximations, we verified that this distribution satisfies the EOS, as can be seen in Eqs.~(\ref{a10}) and (\ref{a11}).  
 \end{itemize}  
\item \textbf{Entropy in the Shell:}  
As shown in Figs.~\ref{entro1} and ~\ref{entroLo1}, the behavior of the entropy in the shell for each distribution is presented as a function of thickness $\epsilon$, where we observed that entropy increases as both the thickness and the minimum length parameter increase. Comparing with the works~\cite{Pradhan:2023wac, Javed:2023nls}, we found that the entropy reaches its maximum value at the shell boundary, a result consistent with our study.  
\end{itemize}

We conclude that the presence of the minimum length in gravastars within a BTZ geometry not only plays a fundamental role in the stability of these objects but also highlights the importance of quantum effects for relativistic understanding on reduced scales. 
\\
\\
\textbf{Data Availability:} No new data are associated with this article.

\acknowledgments

We would like to thank CNPq, CAPES, for partial financial support. 
MAA acknowledge support from CNPq (Grant nos. 306398/2021-4).

\end{document}